\documentclass[]{emulateapj}

\def\Mpc{\ {\rm Mpc}}
\def\kpc{\ {\rm kpc}}

\def\kms{{\ }{\rm km}\,{\rm s}^{-1}}

\def\hdelta{H${\delta}$}

\def\galfit{\texttt{GALFIT}\ }
\def\SE{\texttt{SExtractor}\ }

\def\ref{{\bf ref?}}
\def\J{$J$-}
\def\K{$K_{\rm s}$-}
\def\resp{respectively}

\def\sersic{S\'ersic }
\def\evo{$\to$\ }

\def\Mstel{M_{\ast}}
\def\Msun{M_{\odot}}
\usepackage{amsmath}
\usepackage{MnSymbol}
\usepackage{wasysym}
\usepackage{latexsym}
\usepackage{amssymb}
\usepackage{graphicx}
\usepackage{mathrsfs}
\usepackage[bottom]{footmisc}
\usepackage{color}
\usepackage{hyperref}
\shortauthors{ABRAMSON ET AL.}
\shorttitle{MORPHOLOGICAL EVIDENCE FOR BURST RECYCLING}

\begin{document}

\title{The IMACS Cluster Building Survey: V. Further Evidence for Starburst Recycling from Quantitative Galaxy Morphologies\textnormal{*}}

\slugcomment{Accepted to The Astrophysical Journal}

\author{
Louis E. Abramson\altaffilmark{1,2,$\dagger$}, 
Alan Dressler\altaffilmark{3}, 
Michael D. Gladders\altaffilmark{1,2}, 
Augustus Oemler, Jr.\altaffilmark{3}, 
Bianca M. Poggianti\altaffilmark{4}, 
Andrew Monson\altaffilmark{3}, 
Eric Persson\altaffilmark{3},
and Benedetta Vulcani\altaffilmark{4,5}
}


\begin{abstract}

Using \J\ and \K band imaging obtained as part of the IMACS Cluster Building Survey (ICBS) we measure \sersic indices for 2160 field and cluster galaxies at $0.31 <  z < 0.54$. Using both mass- and magnitude-limited samples, we compare the distributions for spectroscopically determined passive, continuously starforming, starburst, and poststarburst systems and show that previously established spatial and statistical connections between these types extend to their gross morphologies.  Outside of cluster cores, we find close structural ties between starburst and continuously starforming, as well as poststarburst and passive types, {\it but not} between starbursts and poststarbursts.  These results independently support two conclusions presented in a previous ICBS paper (Dressler et al.): 1) most starbursts are the product of a non-disruptive triggering mechanism that is insensitive to global environment, such as minor-mergers; 2) starbursts and poststarbursts generally represent transient phases in the lives of ``normal" starforming and quiescent galaxies, \resp, originating from and returning to these systems in closed ``recycling" loops.  In this picture, spectroscopically identified poststarbursts constitute a minority of all recently terminated starbursts, largely ruling-out the typical starburst as a quenching event in all but the densest environments.\\

\end{abstract}

\keywords{
galaxies: clusters --- 
galaxies: evolution --- 
galaxies: morphologies --- 
galaxies: structure ---
galaxies: starburst ---
galaxies: poststarburst
}

\altaffiltext{*}{Data were obtained using the 6.5-m Magellan Telescopes at Las Campanas Observatory, Chile.}
\altaffiltext{1}{Department of Astronomy \& Astrophysics, University of Chicago, 5640 S Ellis Ave, Chicago, IL 60637}
\altaffiltext{2}{Kavli Institute for Cosmological Physics, University of Chicago, 5640 S Ellis Ave, Chicago, IL 60637}
\altaffiltext{3}{The Observatories of the Carnegie Institution for Science, 813 Santa Barbara St, Pasadena, CA 91101}
\altaffiltext{4}{INAF-Osservatorio Astronomico di Padova, vicolo dell'Osservatorio 5, 35122 Padova, Italy}
\altaffiltext{5}{Kavli Institute for the Physics and Mathematics of the Universe, University of Tokyo, Kashiwa, 277-8582, Japan}
\altaffiltext{$\dagger$}{Corresponding author; \href{mailto:labramson@uchicago.edu}{\tt labramson@uchicago.edu}. Research conducted as a Brinson Fellow at The Observatories of the Carnegie Institution for Science.}


\section{Introduction}

The quiescent galaxy population has grown since $z \sim 1$ in all environments \citep{BO78a, Faber07, Moustakas13}.  Commonly, new passive systems (hereafter PAS) are thought to descend from continuously starforming galaxies (CSF) in which a ``transformational event" depleted or removed cold gas supplies, preventing further star-formation \citep[see][\S5 and references therein]{Renzini06}.  Starbursts (SB) are oft-invoked catalysts for this metamorphosis \citep[e.g.,][]{DresslerGunn83, Couch87, Poggianti99, Quintero04, Hogg06}.

Key to the burst-driven evolutionary scenario is the ``poststarburst" (PSB; a.k.a.\ ``E+A"/``k+a") galaxy population \citep{DresslerGunn83, Zabludoff96, Tran04}.  These objects have spectra exhibiting deep Balmer absorption characteristic of short-lived A stars (signifying recent star-formation) but negligible emission (signifying little current star formation).  Further, they may preferentially occupy a locus in color-magnitude space between the ``blue cloud" of starforming systems and the passive-dominated red sequence \citep[][]{Yan09, Mendel12}.  Because these characteristics make poststarbursts compelling candidates for the CSF--PAS ``missing-link", starbursts emerge as important mechanisms enabling the former's transformation into the latter.  But how important are they?

\begin{deluxetable*}{lccccc}
\centering
\tablecolumns{6}
\tablecaption{ICBS Spectral Type Statistics \& Properties}
\tablehead{
\multicolumn{1}{l}{Type} &
\colhead{Spectoscopic ID} &
\colhead{$N_{\rm tot}$} &
\colhead{$N_{\rm used}$\tablenotemark{a}} &
\colhead{\J band SNR} &
\colhead{\K band SNR} \\
\colhead{} &
\colhead{} &
\colhead{} &
\colhead{$J/K_{\rm s}$} &
\colhead{$25^{\rm th}, 75^{\rm th}$ pctle.} &
\colhead{$25^{\rm th}, 75^{\rm th}$ pctle.}
}
\startdata
Passive 					& PAS	& 457				& 397 (87\%) $/$ 388 (85\%)    & 63, 153 & 68, 130\\
Poststarburst 				& PSB	& 55					& 44 (80\%) $/$ 42 (76\%) 	       & 42, 120 & 50, 119\\
Continuously starforming		& CSF	& 1339				& 1077 (80\%) $/$ 1013 (76\%) & 35, 116  & 50, 115\\
Starburst  					& SB		& 304\tablenotemark{b} 	& 213 (70\%) $/$ 191 (63\%)    & 30, 92   & 43, 95
\enddata
\tablenotetext{a}{Galaxies with $\geq 75\ (50)$  $J$ ($K_{\rm s}$) pixels above 1.5$\sigma_{{\rm sky}}$ with no $r_{\rm e}$, $b/a$, or $n$ fit flags.}
\tablenotetext{b}{233 SBH + 71 SBO}
\label{tbl:stats}
\end{deluxetable*}

In a previous paper from the IMACS Cluster Building Survey \citep[ICBS;][hereafter Paper II]{Dressler13} we asked: What fraction of passive galaxies descended through the CSF \evo SB \evo PSB \evo PAS evolutionary channel?  To address this, we first sought to determine if there were enough intermediate-redshift SBs to account for the growth in the PAS population between then and now.  

Surprisingly, we found far too many in all environments less dense than galaxy cluster cores (i.e., the isolated field, field and cluster-infalling groups, and the supercluster environment at $R_{\rm cl} > 500 \kpc$); assuming the starburst and poststarburst phases have similar lifetimes \citep[$200 \lesssim \tau_{{\rm SB}} \sim \tau_{{\rm PSB}} \lesssim 500\ {\rm Myr}$;][]{Poggianti99, Oemler09} and that all SBs ``quench" appropriately, the number of high-mass passive systems should have approximately doubled within a Gyr after $z \sim 0.4$, far outstripping the observed growth \citep[see e.g.,][]{vanderWel07, Moustakas13}.

Together with our finding of a constant ratio of CSF- to SB-fractions across {\it all} environments, this discrepancy led us to conclude in Paper II that the typical $z \sim 0.4$ starburst does not move into the poststarburst population after a burst subsides.  That is, most starbursts do not go on to exhibit a PSB spectrum and then transform into new passive galaxies.  Instead, as suggested by \citet{Poggianti99}, we posited that SBs generally return to the ``parent" CSF class, remaining in a closed, CSF--SB--CSF recycling loop.

Finding starbursts to be inefficient quenching mechanisms, however, does not mean they are ineffective in terms of accounting for the increase in the passive fraction with time.  To test this, we took the poststarbursts as proxies for those SBs which are quenching, or at least not participating in CSF--SB--CSF recycling.\footnote[6]{We adopt a similar poststarburst definition to that outlined in \citet[][see their \S4]{Poggianti99}, ensuring the majority of these systems are in fact post-burst and not post-truncation galaxies.}  Yet still, PAS overproduction -- now at earlier times -- remained problematic: Given the frequency of $z \sim 1$ PSBs \citep[see][]{Lemaux10} we expect the passive fraction at $z \sim 0.4$ also to be about twice what we observe.  

Combined with a similar constancy of the PAS-to-PSB ratio across all environments, this disagreement suggested that a closed, PAS--PSB--PAS recycling loop must also be active, running parallel to that of the starforming systems.  Hence, many poststarbursts must descend from passive galaxies, not the other way around.

Combining these findings, we constructed the following hypothesis: starforming and passive systems, \resp, typically remain in closed-loop cycles, with base levels of star-formation or general quiescence punctuated by brief periods of starburst activity.  Absent external factors, SBs return to the parent class from which they came with those bursts {\it originating} in passive galaxies briefly appearing as prototypical (spectroscopic) poststarbursts.  

Being unable to seriously alter the relative abundances of their parent types, SB/PSBs are thus not transitional stages in CSF \evo PAS evolution, but transient phases in the lives of these ``normal" systems.  

Given the constancy of the ratio between parent- and burst-type fractions across a wide range of local densities, we proposed minor-mergers involving gas-rich companions as the likely trigger for many intermediate-redshift starbursts \citep[e.g.,][]{Mihos94, Kaviraj09}.  Additional triggers -- such as disk instabilities, tidal encounters, or accretion of cold gas from the inter-galactic medium -- may be operational in the field or in the continuously starforming population.  However, because they occur everywhere \citep{Fakhouri09} and act on all galaxies regardless of host properties \citep[][]{Woods07} minor-mergers are the most compelling candidate for a general mechanism.  

This scenario works well outside of cluster cores.  In these special environments ($R_{\rm cl} \lesssim 500 \kpc$) however, the SB-to-PSB ratio approaches unity and the PAS fraction rises substantially, removing the aforementioned PAS overproduction problem.  Thus, in agreement with many authors, we suggested that other processes dependent on a dense intra-cluster medium (ICM) -- e.g., ram-pressure stripping \citep{GG72} and starvation/strangulation \citep{Larson80, Bekki02, Moran07} -- drive traditional CSF \evo PAS evolution here, both by preventing SBs from returning to a CSF state and actively extinguishing star formation in CSF galaxies.  Because gas disks in low-mass companions are not likely to survive the descent into these extreme environments, most PSBs in cluster cores likely {\it do} reflect the end-states of CSF-derived SBs instead of mergers onto PAS galaxies.

\subsection{An Independent Test}

The conclusions described above were based purely on spectroscopy and photometry.  However, there is a third avenue by which we can examine the role of starbursts in passive galaxy production: galaxy structure.  In this paper, we probe the accuracy of our hypotheses from this independent perspective.  By analyzing the \sersic index distributions of the ICBS spectral types as derived from high-resolution near-infrared (NIR) imaging, we will show that the structural relationships between these systems not only support, but independently suggest, much of the picture we painted in Paper II.

We proceed as follows: Section 2 outlines the ICBS and the data used in this analysis; Section 3 describes our \sersic index measurement routine; Section 4 presents our results; Section 5 our discussion.  We conclude in Section 6.  Throughout, we take $H_0 = 72 \kms \Mpc^{-1}$, $\Omega_m = 0.27$, and $\Omega_{\Lambda} = 0.73$.  All magnitudes are quoted in the 2MASS system unless otherwise indicated.


\section{Data: The IMACS Cluster Building Survey}

The ICBS is a spectrophotometric survey of four 27\arcmin -diameter fields containing five galaxy clusters at $z \sim 0.4$.  Its objective is to characterize the evolution of {\it typical} galaxies across a range of environmental densities at an epoch in which cluster assembly is vigorously ongoing \citep[][]{Kauffmann95, DeLucia04, McBride09, Gao12} and many transformational mechanisms are likely to be active.

Fields were drawn from the Sloan Digital Sky Survey \citep[SDSS --][]{York00} and Red-Sequence Cluster Survey \citep{GladdersRCS05} using the cluster finding technique of \citet{Gladders00}.  Because of its relative insensitivity to relaxation state and our interest in cluster building, we opted for optical selection over gas-dependent techniques (e.g., X-ray selection) to avoid sampling only well-virialized (i.e., built) systems.  The latter are, in some sense, the most extreme environments in the universe, so processes occurring there may not reflect those driving the evolution of average galaxies -- even those in clusters --  at intermediate redshifts.

Details of the optical data comprising the bulk of the survey -- obtained using the Inamori Magellan Areal Camera and Spectrograph \citep[IMACS --][]{DresslerIMACS11} and Low Dispersion Survey Spectrograph III (LDSS3) on the Magellan-Baade \& -Clay telescopes, \resp\ -- can be found in \citet{Oemler13a}.  However, we review key aspects below for convenience.

\subsection{Optical Spectroscopy}

The ICBS is based on over $4800$ spectra ($\lambda / \Delta\lambda \sim 600$) from 42 IMACS and 16 LDSS3 masks (3--4 hr exposures) obtained between 2004 and 2008.  The wide field-of-view of IMACS permits simultaneous coverage out to $R_{\rm cl} \lesssim 5$\ co-moving Mpc, allowing relatively unbiased sampling of the entire (super-)cluster and projected field ecosystems.

The spectroscopic catalog provides uniform, rest-frame spectral coverage from 3700--5200 \AA\ for 2163 objects in the redshift interval $0.31 < z < 0.54$.  These systems are roughly evenly split between five ``metaclusters" (see below; $N = 993$) and the projected, intervening field ($N = 1170$).  Median signal-to-noise ratios (SNR) near 4500 \AA\ range from 30 per 2 \AA\ pixel at $r = 19$ to $\sim 8$ at $r = 22$ (SDSS system).

\begin{deluxetable}{lrcc}
\centering
\setlength{\tabcolsep}{0.03in}
\tablecolumns{4}
\tablecaption{ICBS Environments}
\tablehead{
\colhead{Environment} &
\colhead{$N_{\rm tot}$} &
\multicolumn{2}{c}{$N_{\rm used}$} \\
\colhead{} &
\colhead{} &
\colhead{$J$} &
\colhead{$K_{\rm s}$}
}
\startdata
Field + field groups				& 1164				& 892 (77\%)	&  825 (71\%)\\
Supercluster\tablenotemark{a}		& 471					& 387 (82\%)	&  374 (79\%)\\
Cluster\tablenotemark{b}			& 525				& 447 (85\%)	&  435 (83\%)\\
Cluster core\tablenotemark{c}		& 137					& 119 (87\%)	&  114 (83\%)\\
\textbf{TOTAL}					& 2160\tablenotemark{d}	& 1731 (80\%)	& 1634 (76\%)\\
\textbf{TOTAL NON-CORE}			& 2023				& 1612 (80\%)	& 1520 (75\%)
\enddata
\tablenotetext{a}{Galaxies within $\pm 3000 \kms$ of $\langle z \rangle_{\rm cl}$ and $R_{\rm cl}/R_{200} > 1.5$.}
\tablenotetext{b}{Galaxies within $\pm 3000 \kms$ of $\langle z \rangle_{\rm cl}$ and $R_{\rm cl}/R_{200} \leq 1.5$.}
\tablenotetext{c}{Cluster galaxies with $R_{\rm cl} < 500 \kpc$.}
\tablenotetext{d}{Due to guide-star constraints, 3 objects from the ICBS catalog were not imaged by FourStar.}
\label{tbl:enviro}
\end{deluxetable}

In Paper II these data were used to construct new galaxy spectral types and characterize field and cluster group-scale (sub-)structures.  Below, we consider the four large-scale environments discussed in that paper: (1) the isolated field and field groups; (2) the supercluster and cluster-infalling groups ($R_{\rm cl} / R_{200} > 1.5$); (3) the virialized cluster ($R_{\rm cl} / R_{200} \leq 1.5$); and (4) the cluster core ($R_{\rm cl} \leq 500 \kpc$).  We defer a discussion based on local density to a future paper.

In what follows, we also adopt the spectral type assignments from Paper II.  However, to ease discussion, we combine starbursts identified by EW(\hdelta) (``SBH" in Paper II) with those identified by EW([OII] $\lambda 3727$) (``SBO") into single starburst class (``SB", herein).  The reader should note first that these systems are not necessarily (U)LIRG-like objects, but galaxies whose star-formation rates (SFR) are enhanced by factors of $\sim 3$--10 over the past (few Gyr) average.  Again, we are interested in studying processes affecting typical intermediate-redshift galaxies and therefore characterizing the role average (i.e., moderate) starbursts play in galaxy evolution. Stellar mass ($\Mstel$) and SFR distributions for all types are presented and discussed in Paper II, Figures 2, 3, and \S\S2.2--2.3, \resp.

Also, while both SB classes contain objects at the height of their star formation, many SBHs are probably decaying away from this peak \citep[see][and Paper II, \S\S2.2, 4.6]{Dressler09}.  Our results are qualitatively unchanged, however, if either SB class is examined independently. 

Tables \ref{tbl:stats} and \ref{tbl:enviro} present, \resp, the relevant spectral and environmental definitions and statistics.

\begin{figure}[b!]
\centering
\includegraphics[width = 0.8\columnwidth]{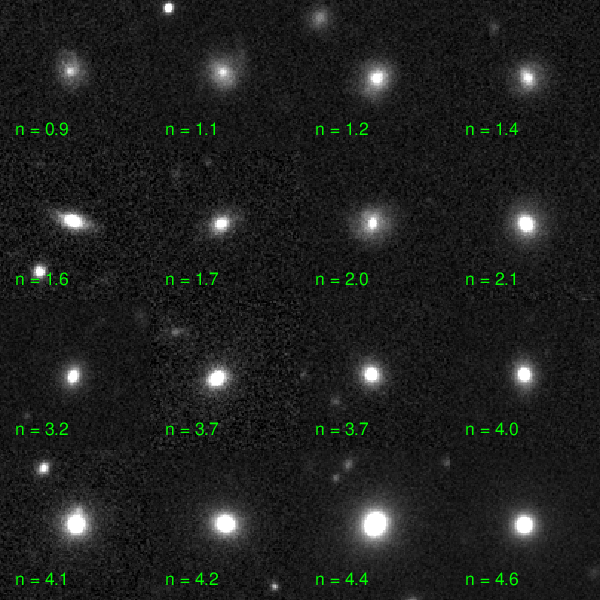}
\caption{\J band source thumbnails arrayed by increasing \sersic index.  Stamp-sizes are $16\arcsec$.  All images are scaled uniformly.}
\label{fig:mosaic}
\end{figure}

\subsection{A Note on Incompleteness}

As discussed in Paper II, the ICBS spectroscopic sample covers approximately 50\% of the photometric catalog down to $r \approx 22.5$.  As it is unbiased with respect to spectral type at the 90--95\% level, no differential incompleteness corrections were implemented in the following analyses.  Significant corrections are necessary if one wishes to discuss mass-dependent phenomena (such as evolution between spectral types) using the magnitude-limited ICBS sample.  However, because we will be largely concerned with characterizing the spectral types individually, we do not apply any mass-incompleteness corrections below.  Instead, we repeat our analyses using a stellar mass-complete sample ($\Mstel \geq 2.5 \times 10^{10}\Msun$) and refer to this when comparing counts across spectral types. We note in advance that results are qualitatively unchanged regardless of which sample is used.

\begin{figure*}[t!]
\centering
\includegraphics[width = \linewidth]{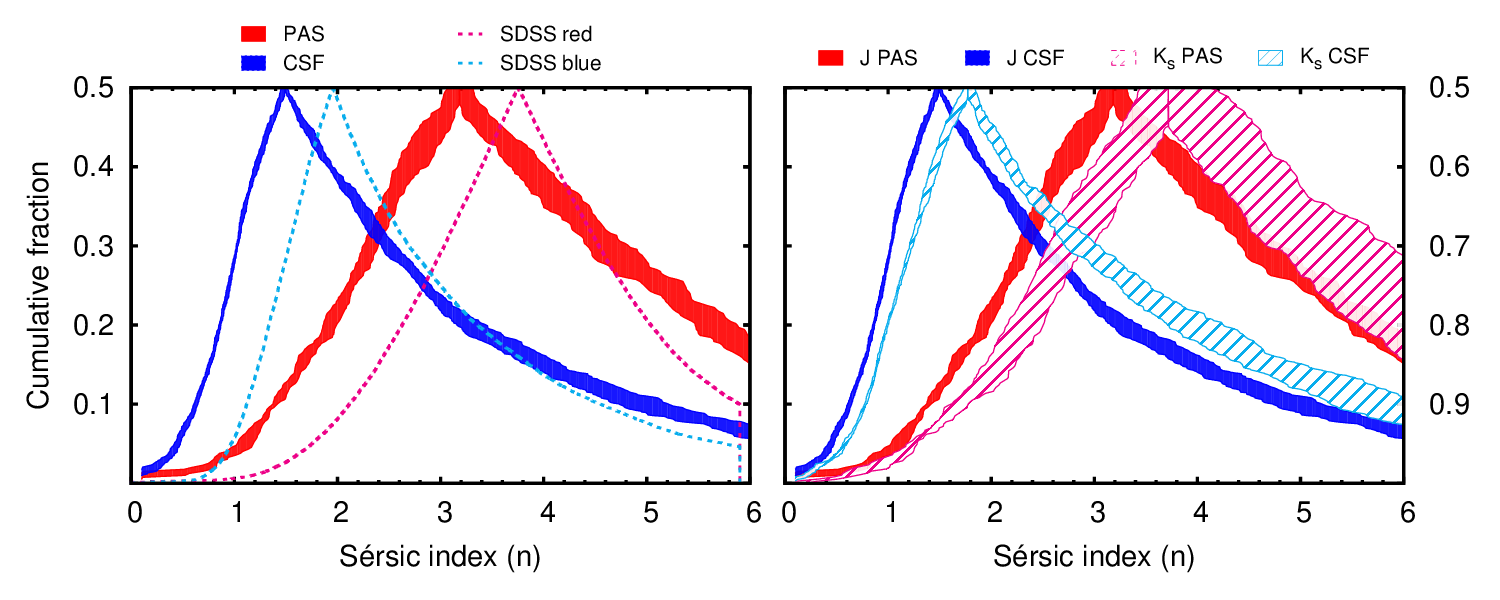}
\vskip -0.25cm
\caption{{\it Left}: Magenta and cyan dashes are, \resp, the folded, cumulative, $z$-band \sersic index distributions (see Section 4) for $\sim 29300$ red and $\sim 14400$ blue galaxies from the SDSS.  Objects are drawn from the group catalog of \citet{Yang07a} with measurements from the VAGC.  Red and blue shaded areas are the $J-$band distributions for $\sim 400$ PAS and $\sim 1100$ CSF galaxies from the ICBS, respectively.  Thickness represents the minimum/maximum values obtained at a given $n$ over six fitting runs.  {\it Right}:  A comparison of our \J\ (solid) and \K band results (striped areas).  Combined with Figure 1, the consistency of these results give us confidence our measurements are (at least statistically) sufficiently robust for the present analysis.}
\label{fig:SDSS_comp}
\end{figure*}

\subsection{NIR Imaging}

Structural parameters are derived from  \J\ and \K band images of the four ICBS fields acquired in November 2011 and February 2012 using the FourStar camera on Magellan-Baade \citep[][]{Persson13}.  At the redshifts considered here, these bands probe the light (dominated by G--K giants) from established stellar populations and are largely insensitive to gas and dust, thus providing almost direct access to the underlying galactic structure which we seek to characterize.  These data will be described in detail in a future paper, but we list here properties relevant to the current analysis.

Each IMACS field was tiled with FourStar in a $3\times3$ mosaic.  Final images ($\sim 0.25\ {\rm deg^2}$) were constructed using A.\ Monson's pipeline, which employs \texttt{SExtractor}, \texttt{SCAMP} and \texttt{SWARP} \citep{BertinSEX, BertinSCAMP, BertinSWARP}, and \texttt{IRAF IMCOMBINE} to astrometer, zeropoint-normalize, distortion-correct, and co-add all pointings across a mosaic.  Astrometric and distortion solutions were computed jointly for \J\ and \K band images, minimizing filter-dependent systematics that might bias morphological measurements.

Typical limiting depths for these images are  $J = 23.3$ and $K_{\rm s} = 21.4$ ($5\sigma$ point-source, 1\farcs0 aperture).  At magnitudes of $J = 21.0$ and $K_{\rm s} = 18.5$ -- encompassing 90--95\% of the spectroscopic targets -- these data yield median SNRs for point sources of 35 (51) in $J$ ($K_{\rm s}$).

Seeing ranged from 0\farcs5--1\farcs0 in $J$ and 0\farcs3--0\farcs9 in $K_{\rm s}$ with median values near 0\farcs55 in both bands.  Since FourStar pixels are 0\farcs16, the point-spread-function (PSF) is well-sampled in all but one image, where it is mildly under-sampled.  As we will show, our results are robust to PSF selection, so no analyses were modified for this field.


\section{Structural Parameter Estimation}

\sersic indexes were measured by fitting single \sersic profiles using \galfit v3.0.4 \citep{PengGALFIT, PengGALFIT3}, with \SE v2.8.6 employed for source detection and ancillary image production.  Software has been developed for automated, batch-mode operation of \galfit -- notably \texttt{GALAPAGOS} \citep{Barden12} -- but flexibility is often sacrificed for speed.  Because our analysis required a spatially variable PSF and the ability to add model components/parameters was considered useful,\footnote[7]{ICBS ground-based imaging spans $grizJK_{\rm s}$.  Though not implemented here, future analyses of the full photometric data set may draw upon these additional capabilities.} we created our own fitting routine. 

For each spectroscopic source, a $200\times200$ kpc stamp was cut from the full, background-subtracted FourStar mosaic and the local gain/exposure time calculated from a coverage map.  A local PSF was then either constructed from an inverse-variance weighted stack of the nearest 10 stars or selected from a library of models.  The latter were created from multi-component \sersic fits to candidate PSF stars and visually inspected to ensure quality.  Below, we discuss results from five runs using this implementation (taking the five nearest PSF models) and one run using the composite, empirical PSF.  In the interest of clarity, all sample statistics are quoted from the ``principal" run using the nearest model PSF.

After PSF selection, \SE produced basic positional, geometric, and photometric data for \galfit input.  The source nearest to the spectroscopic catalog location was defined as the ICBS target (``primary"), but all primaries having no pixel within 1\farcs0 of this fiducial position -- derived from histograms of ICBS--\SE centroid offsets -- were flagged for possible confusion and excluded from later analysis.  These tended to occur where the ICBS source was of low-to-no NIR significance (i.e., rest-frame $B-V \lesssim 0.5$) and resulted in the loss of 57 (136) objects in $J$ ($K_{\rm s}$).

\begin{figure*}[t!]
\centering
\includegraphics[width = \linewidth]{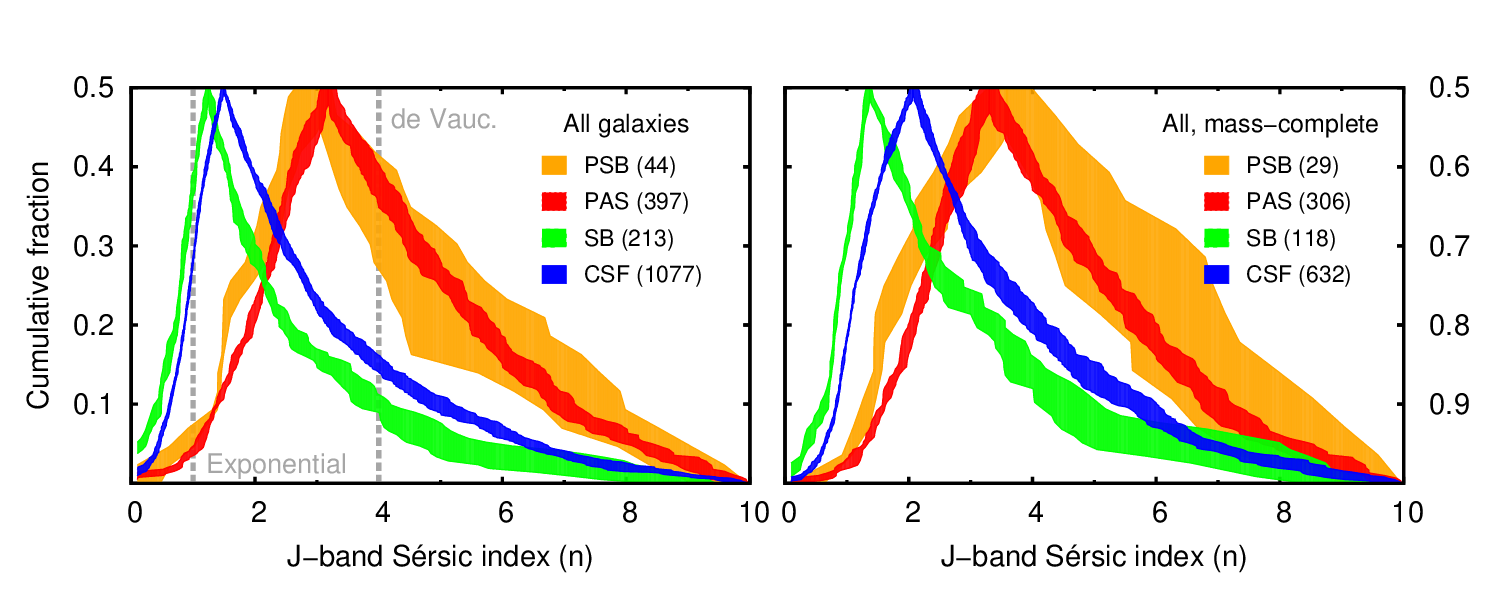}
\vskip -0.25cm
\caption{{\it Left}: \J band \sersic index distributions for all ICBS galaxies.  Quantities in parentheses give $N_{\rm gals}$ in the principal distribution of each type (see Section 3).  Good agreement between PSB (orange) and PAS (red), as well as SB (green) and CSF (blue) distributions is apparent, as is significant discrepancy between those for the SBs and PSBs.  The former tend (in the median) towards the exponential profiles of local spirals, the latter towards the de Vaucouleurs profiles of giant ellipticals. {\it Right}: Distributions for galaxies with $\Mstel \geq 2.5 \times 10^{10}\Msun$, where the ICBS is largely complete for all spectral types.  As these are essentially identical to the magnitude-limited results, mass-incompleteness appears not to be a significant source of bias in our analysis.  That is, the spectral-type structural relationships are not a strong function of $\Mstel$.}
\label{fig:main_result}
\end{figure*}

On-stamp stars were PSF-subtracted before galaxy fitting.  Galaxies with $m - m_p < 2.0\ {\rm and}\ \lvert\overrightharpoon{r - r_p}\rvert \leq 25$ kpc -- where $m_{(p)},\overrightharpoon{r}_{(p)}$\ are magnitudes and transverse positions of (primary) sources -- were fit jointly with the primary.  Pixels outside an ellipse $3.0\ \times$\ \texttt{KRON\_RADIUS} from these ``fit-worthy" galaxies were masked along with non fit-worthy sources and star-subtraction cores.

\galfit was allowed to determine a constant background level for each stamp.  Fixing the sky to zero or a value based on non-source pixel statistics does not affect our results, though it can affect fits for individual sources (see Section 4). 

``Successful" fits were those whose output parameters (half-light radius, $r_{\rm e}$; axis ratio, $b/a$; and \sersic index, $n$) converged away from fitting constraints and were not flagged as unreliable by {\tt GALFIT}.  The first condition essentially imposed a minimum-size criterion which, through experimentation, was found to be $\sim 75$ pixels ($J$) or $\sim 50$ pixels ($K_{\rm s}$) detected at $\geq 1.5\sigma$ above sky fluctuations.  We apply this (somewhat arbitrary) cut, roughly limiting the sample to SNR $\gtrsim 20$, but note that adjusting it either way has a negligible effect on our results.

After excluding mis-identified, poorly fit, and small sources, the final sample contains 1731 (1634) galaxies in $J$ ($K_{\rm s}$) suitable for analysis.  As shown in Table 1, the failure rate is unbiased with respect to spectral type at the $\sim 10\%$ level in $J$ (which we use for the majority of our analyses) so we do not correct for differential failure rate in what follows.


\section{Results}

The model presented in Paper II suggests starbursts and poststarbursts should physically resemble CSF and PAS galaxies, \resp, but not each other.  This is a statement about galaxy morphology and structure.  Ideally, we would test our model in terms of the former as fine-grained details (e.g., the presence of spiral arms or tidal features) place the strongest constraints on formation and transformation mechanisms.  However, for sources at $z \sim 0.4$ true morphologies can be assessed only from space-based imaging and covering the ICBS with the {\it Hubble Space Telescope} ({\it HST}) would require hundreds of pointings, which is practically unfeasible.

Fortunately, high-quality ground-based data (such as we have obtained) is more than adequate to characterize the basic structural properties of our systems.  Since determining even the average diskiness or bulginess of the spectral types would provide a strong test of our hypotheses, we pursue this avenue here.  As is common, we parameterize galactic structure using the \sersic index, $n$ \citep{Sersic68}.

A single, 1200 sec {\it HST} exposure of the core of one of our clusters (SDSS0845 / MACSJ0845.4+0327; Cycle 14 SNAP10491; PI Ebeling) does exist.  We use these data to verify our assessment of the major-merger rate (see Section 5) but do not fit this image because it is in a much bluer bandpass ({\it ACS} F606W), contains $\lesssim 4\%$ of our sources (mostly PASs), and, as discussed above, captures a region where we know our recycling scenario breaks down.  Details of the merger comparison are presented in Appendix \ref{sec:AA}.

\begin{deluxetable*}{lcccccccc}[t!]
\centering
\tablecolumns{9}
\tablecaption{\sersic Index Distribution Statistics, Non-core Galaxies \label{tbl-3}}
\tablehead{
\multicolumn{1}{l}{Type} &
\multicolumn{2}{c}{$\mathscr{L}({\rm \in\ PAS})$ (dex)\tablenotemark{a}} &
\multicolumn{2}{c}{Median $(n)$} &
\multicolumn{2}{c}{$IQR\ (n)$\tablenotemark{b}} &
\multicolumn{2}{c}{$f(n < 2)$\tablenotemark{c}} \\
\colhead{} &
\colhead{$J$} &
\colhead{$K_{\rm s}$} &
\colhead{$J$} &
\colhead{$K_{\rm s}$} &
\colhead{$J$} &
\colhead{$K_{\rm s}$} &
\colhead{$J$} &
\colhead{$K_{\rm s}$}
}
\startdata
PAS	& $\cdots$ 	& $\cdots$  	& $3.1 \pm 0.1$  & $3.7 \pm 0.3$ & $2.9 \pm 0.2$ & $3.1 \pm 0.4$ & $0.23 \pm 0.02$  & $0.15 \pm 0.02$\\
PSB	& $4.1 \pm 0.6$ & $2.9 \pm 0.5$ & $3.2 \pm 0.2 $ & $3.4 \pm 0.3$ & $3.4 \pm 1.0$ & $3.2 \pm 0.3$ & $0.26 \pm 0.03$ & $0.27 \pm 0.01$\\ 
CSF	& $\cdots$ 	& $\cdots$   	& $1.5 \pm 0.1$  & $1.7 \pm 0.1$ & $1.9 \pm 0.1$ & $2.2 \pm 0.1$ & $0.62 \pm 0.01$ & $0.56 \pm 0.01$\\
SB	& $-25 \pm 1$	 & $-26 \pm 1$	& $1.3 \pm 0.1$  & $1.6 \pm 0.1$ & $1.4 \pm 0.1$ & $1.7 \pm 0.1$ & $0.70 \pm 0.01$ & $0.64 \pm 0.01$
\enddata
\tablenotetext{a}{Logarithmic likelihood that a distribution is drawn from the PAS over the CSF parent.  Positive values denote sample is more likely to have come from the PAS class.}
\tablenotetext{b}{Inter-quartile range; $75^{\rm th}\ {\rm minus}\ 25^{\rm th}$ percentile.}
\tablenotetext{c}{Fraction of galaxies with $n < 2$.}
\label{tbl:sersic_stats}
\end{deluxetable*}

\subsection{Fitting Accuracy}

Measuring \sersic indices to high accuracy is difficult \citep[e.g.,][]{Haussler07, Hoyos11, Yoon11}.   For example, across our six runs, uncertainties in individual fits due to background estimation and PSF selection alone range from $\sim 5\%$ at $n = 1$ to $\sim 30\%$ at $n = 4.0{\rm-}4.5$, irrespective of spectral type.  Yet, because we are interested in the relationships between classes of galaxies -- and therefore index {\it distributions} -- we avoid many of the complexities associated with this endeavor.  For our purposes, measured differences will be quantitatively meaningful provided the fits (though uncertain) are unbiased; i.e., provided the spectral types span comparable SNR ranges.  As this is the case (see Table \ref{tbl:stats}), relative comparisons between distributions are reliable.

That said, we believe our measurements are reasonably accurate in an absolute sense.  Though we have not tested our routine on simulated sources, we can assess our fitting accuracy qualitatively and quantitatively in several ways.

Qualitatively, our measured \sersic indices correlate well with visual impressions.  Figure \ref{fig:mosaic} shows \J band cut-outs for some of our higher-SNR sources.  As $n$ increases, systems clearly progress from disk- to bulge-dominated.  Comparisons of ICBS \sersic index histograms (not shown) also appear consistent with optical results from the low-redshift, Wide-field Nearby Galaxy Cluster Survey \citep[][see their Figure 18]{Fasano12}.  From these, we find our PASs to be consistent with a mixed S0/E population and our CSFs to be similar to local spirals, as expected.  

Quantitative assessments are provided in Figure \ref{fig:SDSS_comp}.  In the left panel, we compare $z$-band fits from the NYU Value-Added Galaxy Catalog \citep[VAGC;][]{Blanton05VAGC} for photometrically selected red and blue SDSS galaxies to $J$ measurements (roughly rest-frame $z$) of our PAS and CSF systems.  To approximate the mean ICBS environment, the comparison sources ($z \leq 0.1$) were drawn from the group catalog of \citet{Yang07a}.  

In this and all similar plots below, we show {\it folded} cumulative distributions, i.e., cumulative distributions reflected at their medians.  This (somewhat non-traditional) format gives a good sense of a distribution's width and skew -- as a histogram would -- while avoiding binning.  A complication is that ordinates are reversed at every median.  Hence, the reader must use the left-hand axis to interpret the rising part of each band, but the right-hand axis to interpret the falling part.

There is a systematic shift of $\Delta n \approx -0.5$ and slightly longer high-$n$ tails in the ICBS distributions, but the global similarity of these to the SDSS/VAGC results is apparent.  Given the possible influence of redshift evolution, color versus spectroscopic selection, differences in fitting methods, source resolutions,\footnote{${\rm FWHM_{SDSS}} \approx 1\farcs2 = 1.4 \kpc$ at $z = 0.06$, while ${\rm FWHM_{ICBS}} \approx 0\farcs5 = 2.7 \kpc$ at $z = 0.40$.  Alternately, at those redshifts, 1 kpc spans 2.2 SDSS pixels, but only 1.2 FourStar pixels.  Both types of ``resolution" affect the accuracy of {\tt GALFIT}.} and sample size -- in addition to any real fitting errors -- the correspondence between these results suggests that our measurements are robust for our purposes.  Ultimately, the most relevant aspect to note is the near-identical separation between red and blue distributions in either sample, implying both analyses have comparable power to discriminate between disks and spheroids.

We plot the results of a final test in the right-hand panel of Figure \ref{fig:SDSS_comp}, showing an internal cross-check of our \J\ and \K band results.  Encouragingly, agreement is good.  The small systematic offset between the two measurements could again be due to many factors besides fitting error.  Some of the shift may be physically meaningful (a factor of two in wavelength separates the bandpasses) but the loss of low surface-brightness features (e.g., disks) due to the higher sky backgrounds in $K_{\rm s}$ surely also plays a role.  Regardless, because the offset is essentially uniform, errors contributing to it should not introduce a bias, allowing meaningful comparisons of results obtained independently in either band.  Indeed, as shown in Table \ref{tbl:sersic_stats}, our main results hold across both.

This being the case, given the higher fitting success rate in \J band, we focus on these results below.

\subsection{Spectral-type Distributions}

The left panel of Figure \ref{fig:main_result} presents our first main result: the full \sersic index distributions for all reliably fit ICBS galaxies.  Distribution widths correspond to the minimum/maximum values obtained at a given $n$ over the six fitting runs discussed in Section 3.  We introduce our analytical processes and comparison metrics here before testing the effects global environment and mass completeness have on this result.

The two most obvious characteristics of this plot are also the most important.  First, the SB--CSF and PSB--PAS distributions, \resp, display unambiguous similarities.  Second, there is an equally clear disparity between the SB and PSB distributions; the former is shifted to values characteristic of disks ($n \sim 1$) while the latter is shifted to those of bulge-dominated systems ($n \sim 3$--4).  Although the dissimilarity of the PAS and CSF distributions link these conclusions, there is no {\it a priori} astrophysical reason to expect starbursts to resemble their supposed antecedents (CSFs) but not descendants (PSBs).  Later, we will show that these statements constrain different parts of the recycling model outlined in Section 1.

These (dis-)connections can be quantified in several ways.  Comparing the medians, inter-quartile ranges ($IQR$), and fractions of galaxies with $n < 2$, $f(n < 2)$ -- a generic upper-limit for ``diskiness" \citep[][]{Fisher08} -- confirms visual impressions.  First, the non-starforming types display identical medians ($n \simeq 3.1$) and distribution widths ($IQR \simeq 3.0$), within the scatter of the fitting runs ($\sigma_{{\rm median}} \approx 0.2; \sigma_{{\rm IQR}} \approx 0.7$).  The distributions for the starforming types, though not as consistent, remain quantitatively very similar.  Both have median $n \sim 1.0{\rm -}1.5$ and $f(n < 2) > 60\%$, suggesting that most of these systems are classic exponential disks.  This is to be contrasted with the poststarbursts, which have $f(n < 2) \sim 25\%$.

As a final quantitative comparison, we calculate the relative likelihood, $\mathscr{L}({\rm child \in PAS}$), that the ``child" SB or PSB distributions were drawn from the PAS over the CSF parent:
\begin{equation}
\mathscr{L}({\rm child \in PAS}) \equiv \log\left( \frac{P_{{\rm child,PAS}}}{P_{{\rm child,CSF}}}\right ),
\end{equation}
\noindent where $P$ is the usual probability metric from a two-sided Kolmogorov-Smirnov (KS) test.  This approach uses all the information in the distributions, avoiding binning and parameterization.  

Across our six runs in $J$, we find $\mathscr{L}({\rm PSB \in PAS}) \geq 4.0$ while $\mathscr{L}({\rm SB \in PAS}) < -26$.  In other words, the PSBs have a high likelihood of having come from the PASs (over CSFs) while there is essentially zero probability the SBs were drawn from that parent.  Indeed, the {\it raw} probability $P_{\rm KS}({\rm PSB \in PAS}) \geq 0.38$ for all trials, consistent with the full PSB distribution being drawn from the passive distribution.  Given the limitations of the KS test, however, we do not weigh this fact too heavily.\footnote{For example, the raw probabilities do not directly link SBs to CSFs, but we will argue in Section 5 that this is the only physically acceptable interpretation for most of these systems.}  Regardless, the lack of similarity between starbursts and poststarbursts is clear and real, as is the close structural relationship between the starforming and quiescent types, \resp.  

\subsection{Mass Incompleteness}

So far we have examined the full magnitude-limited ICBS sample ($r \lesssim 22.5$).  Because it gives maximal statistical leverage, this is the best sample to use to characterize the spectral types individually. However, since the types span slightly different mass ranges (see Paper II, Figure 2) it might bias comparisons between them.  To determine if this is the case, we reanalyze a sub-sample containing only galaxies with $\Mstel \geq 2.5 \times 10^{10} \Msun$ (Salpeter IMF; see \citealt{Oemler13a}) corresponding to the $\sim 80\%$ ICBS completeness limit.  The results are plotted in the right-hand panel of Figure 3.  

Clearly, the global trends from the full sample remain largely unchanged; using our log-likelihood statistic, we find $\mathscr{L}({\rm PSB \in PAS)} = 1.6 \pm 0.7$ for the PSBs with $P_{\rm KS}({\rm PSB \in PAS}) > 0.4$ for five of the six runs, but $\mathscr{L}({\rm SB \in PAS)} < -13$.  (Note that some of the change in likelihoods is driven by the reduction in sample size.)  Although the mass-limited CSFs appear slightly bulgier -- perhaps reflecting the correlation of bulginess with mass \citep[][]{Benson07, vanderWel08, Bell12} -- the SBs are still clearly structurally related to these systems and equally clearly distinct from the PSBs.  

Since mass incompleteness does not seem to significantly affect our results, we will continue to use the magnitude-limited sample below.

\begin{figure}
\hskip -0.5cm
\includegraphics[width = 1.1\columnwidth]{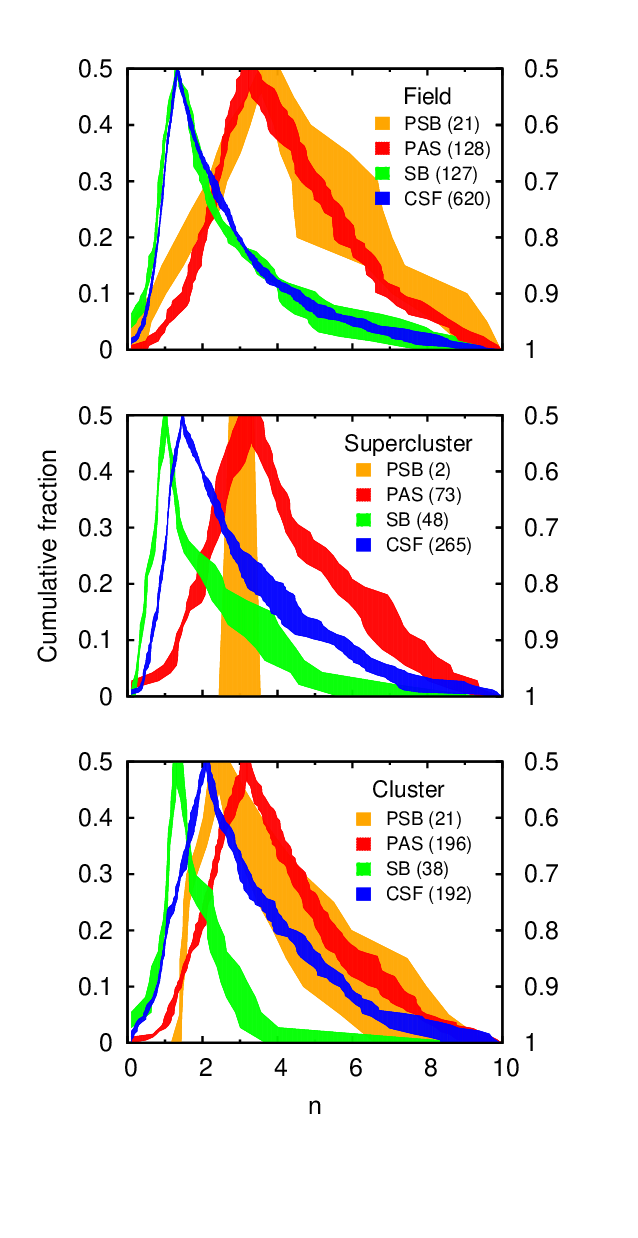}
\vskip -1.7cm
\caption{\sersic distributions for three of the ICBS large-scale environments.  {\it From top}: Field (isolated galaxies and groups); supercluster (infalling galaxies and groups); virialized cluster and cluster core.  In the field, the burst classes are statistically indistinguishable from their respective non-burst ``parents".  In the supercluster, though we have little power to constrain the poststarbursts, the active bursts retain strong similarities to the CSF types.  In the cluster, both burst-types diverge from the non-bursts; PSBs here are diskier than their field counterparts, resembling the CSFs as closely as they do the PASs.  These shifts suggest additional mechanisms may be at work in these dense environments.} 
\label{fig:enviro_results}
\end{figure}

\subsection{Environmental Dependence}

\begin{figure*}[t!]
\centering
\includegraphics[width = \linewidth]{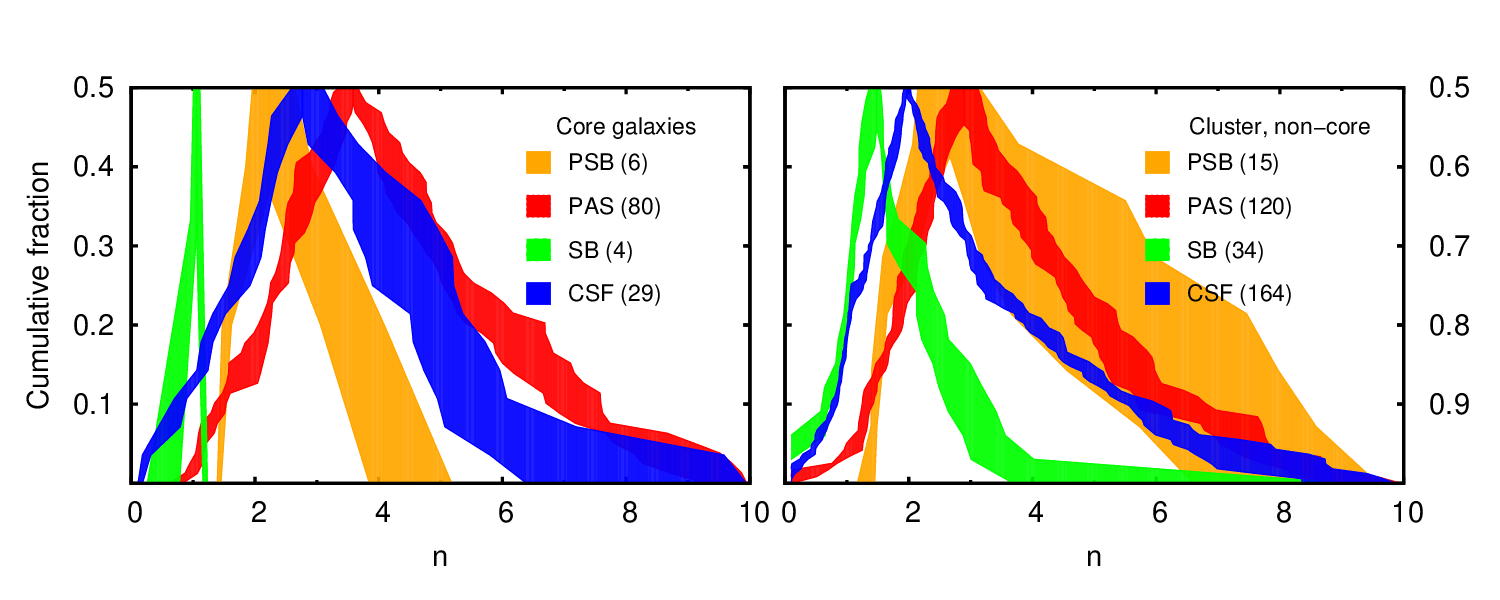}
\vskip -0.25cm
\caption{{\it Left}: \J band \sersic index distributions for cluster core galaxies ($R_{\rm cl} \leq 500$\ kpc).  While the active starbursts in cluster cores are still disky, the parent--daughter relationships between the burst and non-burst classes in this environment are substantially degraded.  Notably, the post-starbursts in the core have \sersic indices between those of the SBs and PASs (or, alternately, close to those of the CSFs).  These facts suggest mechanisms unique to dense environments (e.g., as ram-pressure stripping or starvation) may be causing a ``leak" from CSF to PAS populations through active quenching or starburst-driven evolution.  Population fractions from Paper II also suggest the latter is occurring, here.  Note that, when this population of objects is removed (right panel), the relationship of cluster PSBs to the PASs is strengthened compared to the full cluster sample shown in the bottom panel of Figure 4.}
\label{fig:core_results}
\end{figure*}

All results above were derived using a combination of cluster (core + non-core), supercluster, and field galaxies.  Findings from Paper II suggest such mixing is permissible: the fractional relationships between the spectral types point to recycling being active (if not dominant) everywhere outside of cluster cores.  However, we now test this assumption using the structural data.  We begin by plotting separately the \sersic index distributions for the field, supercluster, and cluster samples in Figure \ref{fig:enviro_results}.  

In the field (top) the picture is identical to the combined result; we find $\mathscr{L}({\rm PSB \in PAS)} = 3.3 \pm 0.6$ for poststarbursts, but $\mathscr{L}({\rm SB \in PAS)} < -12$ for starbursts.  In fact, if anything, the picture here is even clearer: PSBs and SBs have $\langle P_{KS} \rangle \sim 0.5$ of belonging to the PAS or CSF parent, \resp, with no run yielding $P_{KS} < 0.2$.  

In the supercluster (middle) though likelihoods drop significantly (due in part to the smaller number of systems) the same general trends emerge for the starbursts as seen in the field.  These systems exhibit $\mathscr{L}({\rm SB \in PAS)} < -6$, their distribution reflecting the shape of the CSFs to high fidelity though the latter move to slightly higher $n$ overall.  The lack of poststarburst systems prohibits us from constraining their relationship to the PASs, but we note that the two well-fit PSBs in this environment have \sersic indices falling precisely at the median value of the PAS distribution ($n \simeq 3$).  

However, in the cluster proper (bottom) the picture changes.  First, it is clear that CSFs in this environment are considerably more bulge-dominated than those in the field.  Their subtle departure from the starbursts in the supercluster is also exacerbated.  This displacement may simply be a manifestation of the well-known morphology--density relation \citep{Dressler80, Dressler97, Postman05} though interestingly it is not seen in the PAS population.  Conversely, cluster PSBs appear to be diskier than their field counterparts, moving closer to the starforming systems.  This shift is reflected by the KS statistics: with $0.1 < \mathscr{L}({\rm PSB \in PAS)} < 1.1$, PSBs appear only marginally more likely to have come from the PASs over the CSFs.  

Yet, PAS--PSB--PAS recycling may still be active in the cluster environment.  If the dense ICM of the cluster {\it core} is providing additional processing as we expect, core galaxies may be significantly biasing otherwise similar trends away from those of the field and supercluster.

We test this in Figure \ref{fig:core_results}, plotting the distributions for a ``core-only" sample (left) and the cluster with those galaxies removed (right).  Although statistics are limited, much of the shift to diskier PSBs indeed appears to be driven by systems in the innermost 500 kpc of the cluster (modulo projection effects) where ram-pressure or tidal stripping may be playing large roles.  

Comparing the core-excised sample to the full cluster sample (Figure \ref{fig:enviro_results}, bottom) reveals the gap at low-$n$ between the PSBs and PASs to have largely disappeared.  Though, to the eye, there may still be some ambiguity between the non-core PSBs and CSFs at low-$n$, the KS metric reveals the former now to be 63 times more likely to have come from the PASs on average, up from $\sim 3$ in the full cluster sample.  The raw KS probability $P_{KS}({\rm PSB \in PAS})$ is also always greater than 0.2, while in four of the six runs $P_{KS}({\rm PSB \in CSF})$ is less than 1 percent.

We note that these likelihoods represent conservative bounds to the true probabilities since the core-excised sample almost certainly includes ``overshoot"/``backsplash" galaxies \citep{Balogh00, Moore04, Bahe12}, i.e., systems which have been processed by the core but now lie at larger radii.

Given the trend of KS results, it seems that the {\it unprocessed} cluster population likely exhibits the same structural connections as those in the field and supercluster environments. An examination of galaxies in field and cluster-infalling groups also yields results entirely consistent with those of the field and supercluster.  Thus -- as suggested in Paper II -- it indeed seems that there are only two significant environments in terms of the relationships between the spectral types: the highest-density regions of the universe, and everywhere else.

\begin{figure}[t]
\hskip -0.7cm
\includegraphics[width = 1.2\columnwidth]{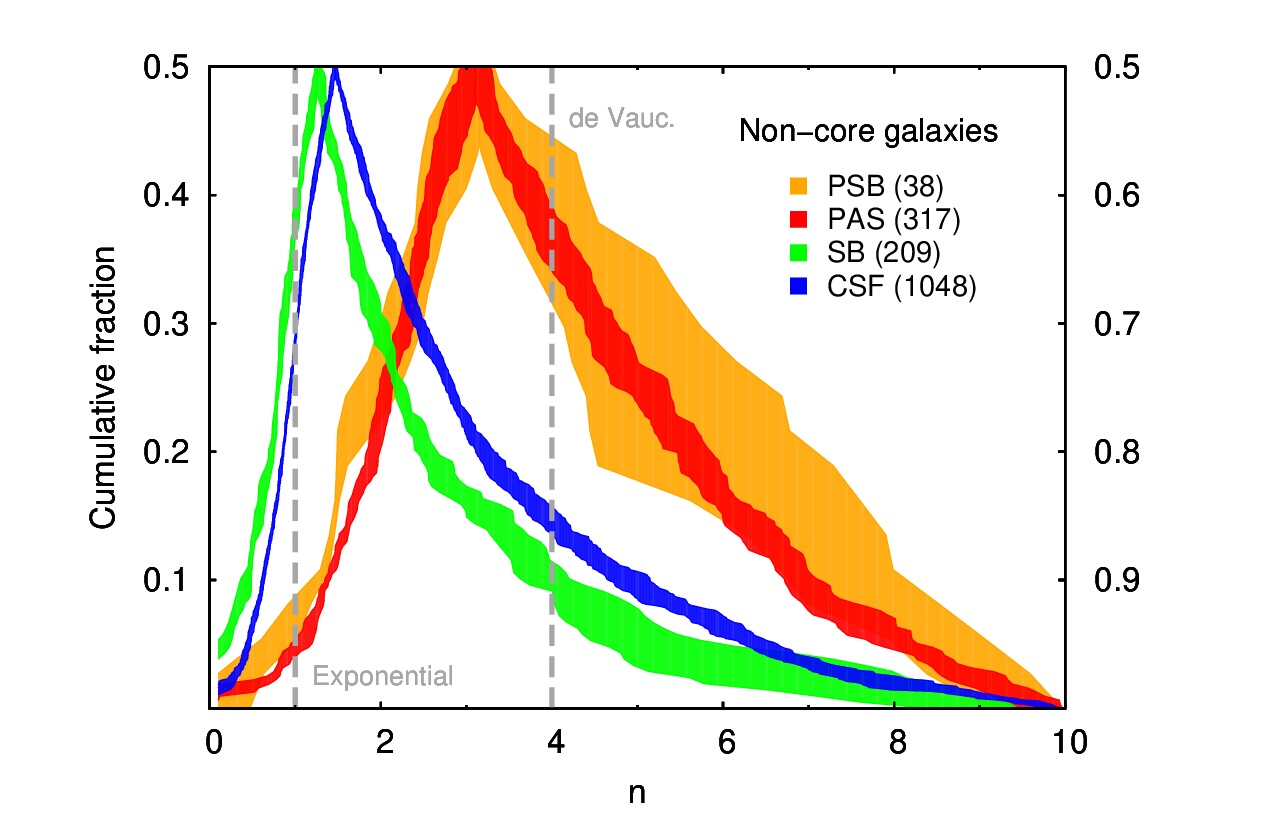}
\vskip -0.2cm
\caption{Final \sersic distributions for all galaxies residing outside of cluster cores.  Properties are listed in Table \ref{tbl:sersic_stats}.}
\label{fig:final_results}
\end{figure}

If galaxies living ``everywhere else" (i.e., the overwhelming majority of systems) are examined, one obtains the distributions plotted in Figure \ref{fig:final_results}.  Here, the spectral type relationships are entirely unambiguous.  Statistics -- medians, inter-quartile ranges, $n < 2$ fractions, and $\mathscr{L}({\rm child \in PAS)}$ values -- describing these ``non-core" distributions are presented in Table \ref{tbl:sersic_stats}.  Unless otherwise specified, the discussion in the next section will refer to this sample.


\section{Discussion}

As shown in the previous section, the structural connections between the spectral types are the same as those exhibited by their population fractions: the SB and CSF as well as PSB and PAS types resemble each other closely, but the active- and post-starbursts are highly dissimilar.  However, while necessary, showing that starbursts are disk-dominated and poststarbursts are bulge-dominated is not sufficient to demonstrate that the closed recycling loops we posited are in fact operational.  Indeed, many others -- using both 1D and 2D fitting techniques -- have found low-redshift poststarbursts to be comparably bulge-dominated to passive systems \citep[see e.g.,][]{Quintero04, Balogh05, Yang08, Mendel12, Bell12} but used this to support the traditional CSF \evo SB \evo PSB \evo PAS quenching scenario we believe to be sub-dominant. We now show that these structural relationships are consistent with -- and indeed independently suggestive of -- the recycling scenario we presented in Paper II.

We turn first to the starforming systems.  It is clear that the starbursts are, in general, the diskiest galaxies in all environments.  This fact is key: it implies that whatever is triggering the majority of bursts must be gentle.  Whether they actively destroy disks or merely build bulges, violent interactions would wash-out the strong clustering around $n = 1$ displayed by these systems.  That this is not the case suggests most SBs cannot evolve into the bulge-dominated PSBs; they are not undergoing the necessary structural transformation.  It therefore seems that these systems {\it have no choice} but to return to the CSF population after their current episode of enhanced star-formation subsides.  (This is true even if, in some environments, the latter class is slightly ``bulgier" than the SBs, on average.)  Thus, the starforming recycling loop we proposed earlier emerges naturally from the structural data as well.

But what of the quiescent galaxies?  As mentioned in Section 1, because SBs so outnumber PSBs, showing that most CSF-derived SBs do not evolve into PSBs does not imply that most PSBs do not descend from CSF-derived SBs.  To test the second statement, we again take the PSBs as proxies for all starbursts not involved in CSF--SB--CSF recycling.  The bulginess of these systems then suggests one of two things: (1) PSBs represent the subset of CSF-derived starbursts which {\it have} undergone major transformational events (i.e., major-mergers) and are now quenching; (2) PSBs originate in systems which are bulge-dominated {\it ab initio}.

Unfortunately, the structural similarity between the PSB--PAS classes alone is not enough to clarify this ambiguity.  As they are bulge-dominated, pressure-support is significant in the non-starforming systems.  Therefore, we would not expect the signature of any transformational mechanism to be easily detectable.  Hence, given only a strong ``family resemblance" we cannot immediately differentiate between the two cases outlined above. However, we can make progress by attacking the problem from the opposite end, asking:  Are there sufficient SBs involved in major-mergers to account for the PSB population?  If so, PSB \evo PAS evolution could explain the structural trends.  If not, our recycling scenario would be favored.

To address this question, L.A. performed a visual inspection of the \J band data, looking for galaxies that displayed clear signs of interactions with similar-sized neighbors without reference to their spectral type.\footnote{Results from {\it HST} data are inconclusive.  Cluster surveys \citep[e.g.,][]{Dressler99, Tran03, Poggianti09} find PSBs to be generally pristine late- or early-type disks.  Field studies \citep[e.g.,][]{Tran04, Yang08} find many to display morphological irregularities indicative of recent mergers.}

\begin{figure*}[t!]
\centering
\includegraphics[width = 1.5\columnwidth]{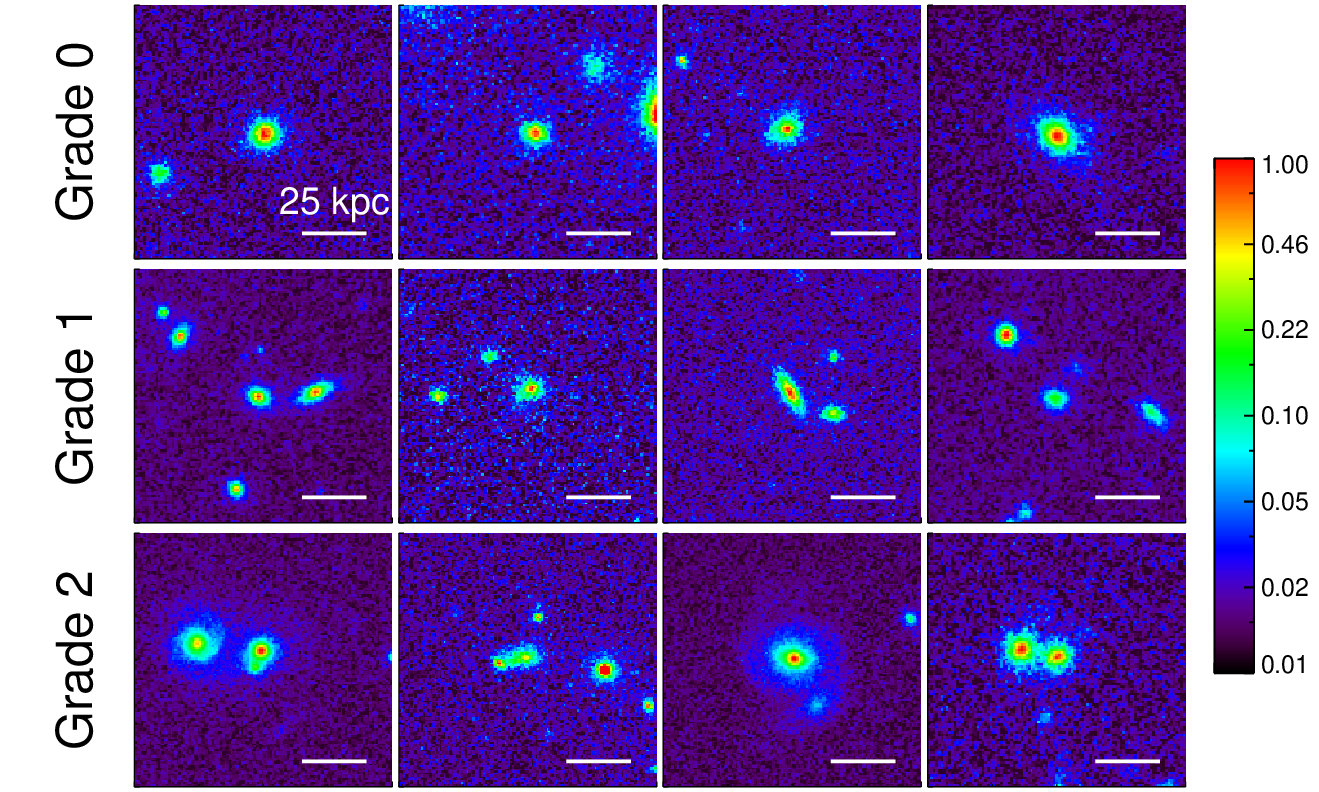}
\caption{Representative \J band cut-outs for Grade 0 (non-merging), Grade 1 (possibly merging), and Grade 2 (``definitely" merging) galaxies.  Each $100\times100 \kpc$ stamp is centered on the fiducial ICBS spectroscopic target location.  We are concerned only with {\it major-mergers}, so a high grade is not given to perhaps ongoing mergers where the galaxies are not of comparable luminosity.  Scaling is {\it asinh} relative to maximum.}
\label{fig:mergers}
\end{figure*}

Each object was graded from $100\times100 \kpc$ cut-outs on a scale of 0 to 2 (0 = no evidence of merging, 2 = definite merger in progress) with a subset graded twice after a random rotation and/or reflection.  Of these objects, about 10\% moved from grade 1 (possible merger; close/small companion or tidal feature) to 2 (obvious disruption from neighbor, large tidal tails, ``train-wreck" appearance) upon second viewing. This ``upgrade"rate agrees well with that obtained by comparing ground- to space-based grades using the single {\it HST ACS} image in the ICBS footprint (see Section 4 and Appendix \ref{sec:AA}).  Hence, we include as ``confirmed" major-mergers all grade 2 plus an additional 10\% of the number of grade 1 systems. Representative cut-outs are presented in Figure \ref{fig:mergers}.

We did not quantify mass-ratios for these possible mergers.  However, the comparable sizes and luminosities of the galaxies involved suggest that the usual 1:3--1:1 definition for ``major" mergers applies (see Figure \ref{fig:mergers}).

In our combined, mass-limited, non-core sample (drawn from the full ICBS catalog, not the subset of successfully fit galaxies) L.A. found $3_{-1}^{+2}\%$ (PAS) to $6_{-1}^{+2}\%$ (SB) to be involved in major-mergers. Errors reflect 68\% confidence assuming a ``beta distribution" for merger probabilities \citep{Cameron11}.  This result agrees well with that of \citet{Bell06} -- who find $5\%\pm1\%$ of all galaxies at $0.4 < z < 0.8$ to be merging using the same stellar-mass limit we apply ($\Mstel \geq 2.5 \times 10^{10} \Msun$) -- and also \citet{Williams11} -- who find a $6\% \pm 1\%$ merger fraction for galaxies at $0.4 < z < 2.0$ with $\Mstel \geq 3.2 \times 10^{10} \Msun$.

Indeed, an independent inspection by A.O. yielded merger fractions 40\% {\it lower} than those quoted above, in agreement with a previous {\it HST} study of $z \sim 0.4$ groups by \citet[][see their Table 1]{Wilman09}.  While they may therefore be closer to upper-bounds, since we wish to constrain the maximal impact of major-mergers, we discuss the more-generous estimates in what follows.

For the SB class, this fraction corresponds to 11 systems.\footnote{The mass-limit for these systems is extended down to $\Mstel \geq 1.7 \times 10^{10} \Msun$ to account for up-scattering of 1:3 mergers into the full mass-complete sample.}  There are 26 PSBs in the sample.  At face value then, the number of major-merging active starbursts can account for perhaps a third to half -- but not most -- of the poststarbursts. Notably, this is very similar to the offset between our PSB fraction ($\sim 2\%$) and that estimated to be due to major-mergers at $z \sim 0.4$ from \citet{Snyder11}.  Starburst recycling must therefore be very active (if not dominant) as we suggest in Paper II.

Although, as discussed above, we believe it is fairly accurate, we acknowledge that this is an imprecise estimate.  Two systematics drive this uncertainty: merger identification and remnant properties.  We constrain the effects of these issues, showing that they should not invalidate the scenario we have outlined above,  in the following sections.

\subsection{Merger Identification Uncertainties}

Regarding the issue of identification, there are two concerns: (1) merger features (e.g., tidal tails) may not be obvious at all merger stages; (2) mergers might occur on timescales much shorter than those over which spectral indicators change.  If either is the case, we would underestimate the true number of merging starbursts.

With respect to the latter concern, while the range of theoretical values for major-merger visibility timescales is large -- radial separations and morphological disturbances depend on the myriad configurations and properties of the galaxies involved -- it seems safe to say that $\sim 0.5-2.0\ {\rm Gyr}$ are reliable bounds \citep[e.g.,][]{Lotz08, Conselice09, Lotz10a, Lotz10b}.  Considering our data span approximately 2 Gyr and SB indicators are sensitive to timescales $\gtrsim 200$ Myr, neither our data set nor spectral categorization should significantly under-sample the merger rate.  That is, our data provide a sufficiently long baseline to capture most of a merger and our spectral definitions respond quickly enough to ensure that most merging starbursts fall in the SB class.

Misidentification of merging systems as non-mergers -- because, for example, the galaxies are at large separations, have just coalesced, or do not display disturbed morphologies -- is more problematic.  Turning to \citet[][see their Figures 4 and 11]{Lotz08} and examining their ``Sbc" model (slightly more massive than our average SB but consistent with its diskiness) we find that, although (projected) pair separations of $> 50 \kpc$ are expected to last for perhaps 20\% of a merger, high SFRs can persist after coalescence for almost a gigayear.  Thus, missing mergers due to partners falling-off inspection stamps should not be a significant problem, but grading ``just-merged" systems as non-mergers might be.  

Fortunately, asymmetry metrics can be high during this period.  Although a comparison of visual to quantitative metrics is not ideal, this suggests we should have captured many coalesced objects; highly-disturbed ``isolated" systems would receive high merger grades.  

Hence, we believe the dominant source of identification uncertainty is likely to be the fraction of a merger over which prominent asymmetries are visible, which is about a third to half.  In the maximal case, then, we might have underestimated the number of major-merging starbursts by a factor of about three. 

However, though identification uncertainties {\it might} in this way permit the SB merger-fraction to account for the number of PSBs, uncertainties in the efficacy of such events in creating bulge-dominated, quiescent remnants appear to run in the opposite direction.

\subsection{Uncertainties in Remnant Properties}

Interestingly, {\it none} of the ``Sbc" mergers from \citet{Lotz08} terminate in non-starforming systems. Although these authors do not employ AGN feedback in their simulations, the effectiveness of this process in stifling star formation or leading to poststarburst remnants remains questionable \citep{Brown09, Wild09, DeBuhr10, DeBuhr11, Snyder11}.  Furhter, according to \citet[][see their Figure 15]{Hopkins10}, assuming most of our disky SBs are indeed CSF-derived or fall into their ``gas-rich" category, bulge formation is preferentially {\it suppressed} in mergers of these systems compared to those involving bulge-dominated or gas-poor galaxies.  Combined, these effects might substantially reduce the number of major-merging SBs that could result in both bulge-dominated and quiescent remnants, i.e., systems which actually resemble our PSBs.

One could argue that the companion of a merging SB -- which may not be captured in the spectroscopic catalog -- might be bulge-dominated or gas-poor and thus that the remnant would efficiently grow a bulge.  However, mergers involving local red and blue galaxies -- a proxy for this scenario -- have been estimated by \citet{Chou12} to be about ten times less common than mergers involving two blue galaxies.  Therefore, though the analogy is not perfect, the probability that most of our SB mergers involve a gas-poor companion appears low.

\begin{figure}[t!]
\hskip -0.7cm
\includegraphics[width = 1.1\columnwidth]{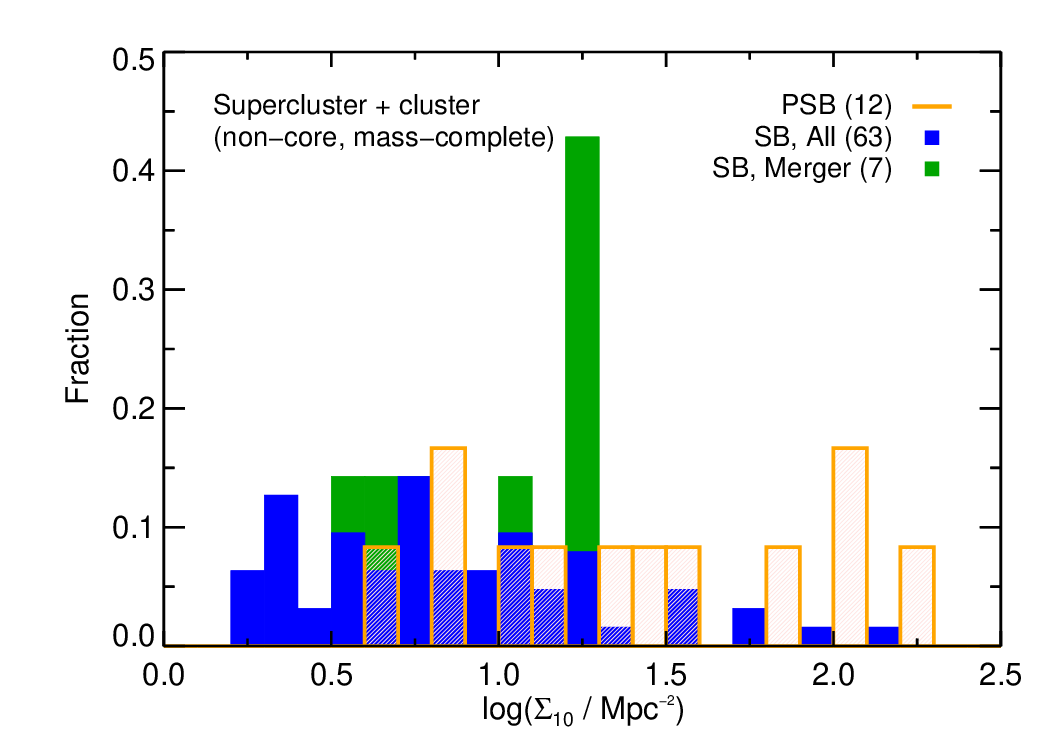}
\vskip -0.2cm
\caption{Surface densities (measured to the ten nearest-neighbors) of the environments of cluster PSB and SB systems.  SBs plausibly involved in major-mergers (green histogram; see text Section 5.1) sample the underlying starburst population (blue histogram) in an unbiased fashion while PSBs (orange shaded histogram) live preferentially in higher-density regions.}
\label{fig:merge_hist}
\end{figure}

There may also be a more general constraint to consider.  Both numerical \citep[e.g.,][]{Lotz08, Lotz10b} and observational studies \citep[e.g.,][]{Poggianti99} suggest that bulge-building (in mergers or otherwise) is delayed with respect to the cessation of starburst activity.  If the delay is significant and a large portion of poststarbursts come from CSF-derived SBs (likely to be disk-dominated) then a substantial fraction of PSBs should be disky.  This is clearly not the case (see Table \ref{tbl:sersic_stats}).

In sum, uncertainties in remnant properties and the effects of delayed morphological versus spectroscopic transformation probably (greatly) suppress any boosts mis-identification issues give to the number of major-merging SBs.  So, although we cannot rule-out CSF-derived, major-merging SBs as progenitors of a sizable fraction of our PSBs it seems unlikely that they are responsible for most of this population given the factor of $\sim 2$--3 baseline short-fall in numbers.

\subsection{Environments of Major-mergers: An Additional Constraint}

One final piece of evidence speaks against a CSF-derived, major-merger driven origin for most poststarbursts: the cluster galaxy correlation function.  As shown in Paper II (see Figures 4 and 6, therein) the active- and post-starbursts (at least in clusters) have very different spatial distributions.  Perhaps unsurprisingly, the SBs track CSFs, appearing relatively uniformly over the faces of our clusters, while the PSBs track the PASs, remaining centrally concentrated.  (Recall the paucity of PSBs in the supercluster sample, above.)  Therefore, even if we have somehow massively underestimated the number of mergers or if all of the merging SBs will in fact become quiescent spheroids (neither of which do we think is true) they would still have to rearrange themselves {\it spatially} in order to account for all of the PSBs.  

To test this, we relaxed our definition of ``confirmed" major-mergers to include all (mass-complete) SBs with grades $> 0$ and compared the local environments of these galaxies to those of the PSBs.  (There are too few of these systems to adequately constrain their correlation function.)  The results -- using the surface density of a galaxy's ten nearest-neighbors to parameterize ``local environment" -- are plotted in Figure \ref{fig:merge_hist}.\footnote{This analysis for field galaxies reveals no significant difference between the density distributions of any of our spectral types.  This is likely because the spectroscopic catalog samples the volume in this environment too sparsely to obtain a meaningful measurement.}  

\begin{figure}[t!]
\hskip-0.5cm
\includegraphics[width = 1.1\columnwidth]{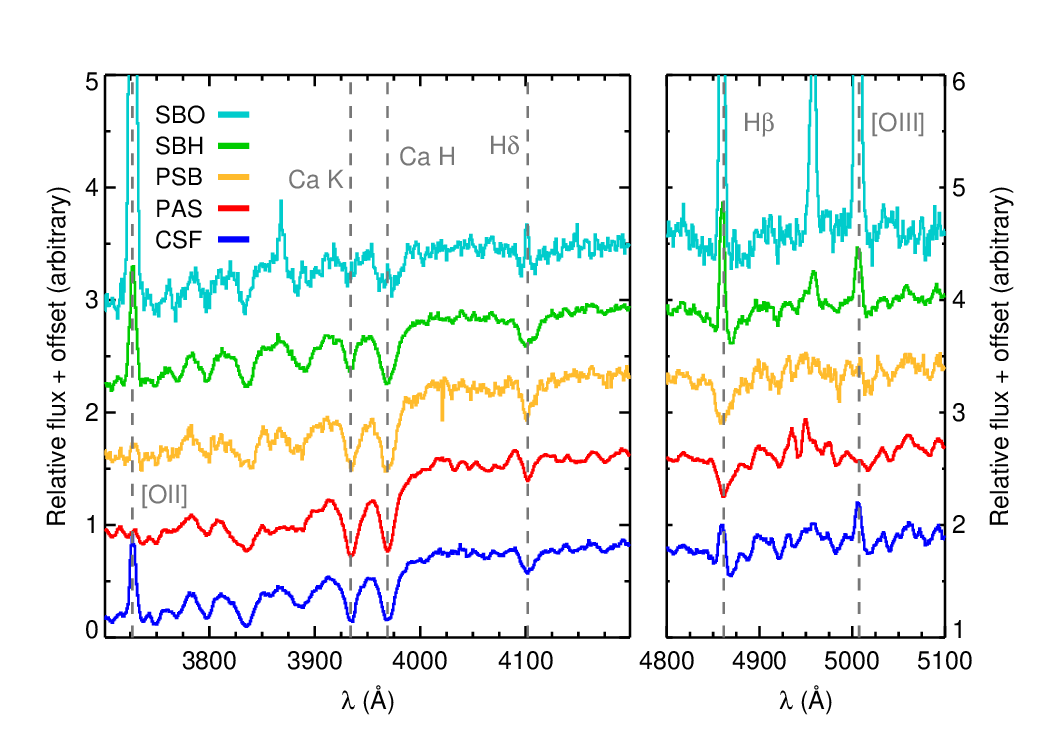}
\vskip -0.3cm
\caption{Mean composite spectra of the ICBS spectral types normalized to the CSF continuum surrounding \hdelta\ and [OIII].  Note that many SBHs -- which comprise 77\% of the SBs considered here -- are {\it Spitzer}-$24\ \micron$ sources and [OII] emission may therefore be strongly extinguished \citep[see also][]{Kocevski11}.  A considerable number of these systems may also be decaying starbursts  \citep[see][and Paper II]{Dressler09}.  Although spectroscopically the SBH--PSB ties are strong -- the former appearing as PSBs with emission -- structurally they are quite divergent, suggesting evolutionary connections are weak.}
\label{fig:spectra}
\end{figure}

From these histograms we see that (plausibly) merging SBs live in regions with densities similar to those of the rest of the starburst population.  The PSBs, however, clearly do not.  Indeed, approximately half of the latter are found at densities where there are no major-merging SBs (though still some non-merging SBs). 

A larger sample is needed to draw definitive conclusions, but these data suggest that many major-merging, CSF-derived SBs do not have the proper spatial distribution to be the progenitors of our poststarbursts, even if the aforementioned uncertainties in merger fractions, etc., might allow them to account for the total number of these systems.  Further, tests show this result to be independent of SFR, $\Mstel$, and $n$.  Hence, if anything, this plot suggests that we may have {\it overestimated} the impact of mergers by about a factor of two; only half of our PSBs live in similar environments to the merging SBs.  This further reduces the impact of possible mis-identifications discussed above.

These results might appear to disagree with those of \citet{Kocevski11}, who, in a study of $z \sim 0.9$ clusters, found a larger fraction of SBs ($\sim 25\%$) to be merging and to prefer high-density environments.  However, this analysis focused mainly on LIRGs, objects with substantially higher $24\ \mu$m fluxes and SFRs than our typical SB (see Section 2.1 and Paper II).  Combined with possible evolution in the SB population between $z \sim 0.9$ and $z \sim 0.4$ ($\Delta t \approx 3$ Gyr), the fact that our sample contains very few LIRGs may explain why we do not see a similarly enhanced major-merger rate or preference for high-density environments in our SBs.

It is of course possible that a conspiracy is at work, i.e., that PSBs reflect the most efficient bulge-forming mergers and we are missing those SBs which recently underwent a major-merger in the denser regions of our superclusters.  

However, a more straightforward interpretation of the data -- which show a close resemblance of the poststarbursts to an existing population of appropriately bulge-dominated galaxies with an inherently similar spatial correlation function -- is that a large fraction of these galaxies are simply born from the passive population.  If so, a non-starforming PAS--PSB--PAS recycling loop is almost certainly operating in parallel to the starforming loop described, above.

We stress that, from a purely spectroscopic standpoint, this result is surprising.  Figure \ref{fig:spectra} shows composite spectra of the ICBS spectral types.  The resemblance between our SBs and PSBs (especially in the depth of \hdelta, higher-order Balmer lines, and the relative strengths of Ca H and K) is apparent, the latter appearing essentially as emission-less versions of the former.  Hence, given only these data, the evolutionary scenario has great appeal!  It is only when information is combined across multiple domains -- photometric, spectroscopic, and morphological -- that persuasive alternative interpretations emerge.

\subsection{The Starburst Mechanism}

A range of plausible (if not operational) triggering mechanisms are consistent with the SB--CSF structural relationship we find.  These include tidal disruptions, intrinsic disk instabilities, and gas accretion, as well as minor-merger.  However, minor-mergers appear to be the most likely candidate for producing the PAS-derived starbursts that lead to most PSBs.  The reason is simply that passive galaxies lack large gaseous disks and are known to possess hot halos which would stifle cold-mode IGM accretion \citep[][]{Forman85, Mulchaey10}.  Further, taking numbers from \citet{Lotz11} and \citet{Newman12a}, \resp, minor-merger rates of $\sim 3$ times the major-merger rate and close-companion fractions of $\sim 13-18\%$ are both about the right size to allow these interactions to explain the ${\rm \sim 1:10}$ PSB:PAS ratio we find.  While we cannot definitively say what fraction of {\it all} SBs (CSF- and PAS-derived) are the result of minor-mergers, it is reasonable to suspect that these events occur in a similar fashion across PAS/CSF hosts \citep{Woods07} suggesting that such interactions could be a significant-to-dominant source for intermediate-redshift starbursts.

\subsection{Cluster Cores: an Aside}

As noted in Section 1, the recycling scenario discussed above is expected to break down in cluster cores.  Here, the SB:PSB ratio climbs to $\sim 1$ and the PAS fraction rises rapidly at the expense of the CSFs, implying CSF \evo PAS evolution is active.  This is not surprising: the dense ICM in these regions should prevent infalling SBs from rejoining the CSF population (breaking the starforming recycling loop) and -- as has been known for decades -- actively quench CSF galaxies through, e.g., ram-pressure stripping.  While starbursts are thus only incidentally connected to PAS build-up -- the global extinguishing of star formation affects the more-numerous CSFs as well as the SBs -- we do expect them to be the dominant source of core PSBs: no low-mass, gas-rich systems should survive long enough in these regions to accrete onto PASs.

Though projection effects and small sample size prohibit drawing definitive conclusions, the plot of cluster core \sersic index distributions in Figure 5 (left) suggests something consistent with this scenario is taking place.  Here, the PSBs are seen to lose their high-$n$ tail \citep[as noted previously by the MORPHS collaboration,][]{Dressler99} and depart significantly from the PASs, falling squarely between these and the still-disky SBs.  This is the signal we would expect if core SBs and PSBs reflect larger-radius accretions onto disky CSFs and their subsequent ICM-driven quenching.  However, the similarity of core PSBs to core CSFs also suggests that some of these systems may reflect ICM-triggered bursts \citep{Bekki03} or the most extreme examples of post-truncation galaxies.


\section{Conclusion}

Using large, mass- and flux-limited samples from the IMACS Cluster Building Survey, we measured the \sersic indices for intermediate-redshift galaxies of all spectral types from high-quality ground-based NIR imaging.  Our results support the existence of a starburst recycling scenario presented in \citet{Dressler13} operating in environments from the isolated field to rich clusters.  We find:

\begin{itemize}
\item Little-to-no structural similarity between active and poststarburst systems outside of cluster cores, indicating that CSF \evo SB \evo PSB \evo PAS evolution is weak in almost every environment modulo some uncertainty about the contribution of major-mergers;
\item Strong structural ties between SB--CSF and PSB--PAS classes, suggesting that most (post-)starbursts are transient ``blips" in the lives of ordinary starforming and quiescent galaxies and do not represent stages in galaxy ``quenching";
\item The star-forming/-bursting systems to be disky, everywhere.  Taken with the above relationships, this independently suggests that a gentle mechanism (likely minor-mergers) is responsible for the production of the typical intermediate-redshift starburst;
\item Evidence that this picture may reverse in cluster cores, with environmentally specific agents providing a channel for CSF \evo SB \evo PSB \evo PAS evolution.
\end{itemize}


\section*{Acknowledgements}
	
LA thanks Drs.\ Daniel Kelson and Chen Peng as well as Song Huang for generously sharing their time and insight.  He also thanks Gabriel Prieto, Geraldo Valladares, and the rest of the LCO staff for their patience in supporting and skill in executing the observations analyzed here.  AD and AO acknowledge the support of the NSF grant AST-0407343.  All the authors thank the referee for providing useful suggestion and NASA for its support through NASA-JPL 1310394.  BV and BP acknowledge financial support from ASI contract I/016/07/0 and ASI-INAF I/009/10/0. MDG thanks the Research Corporation for support of this work through a Cottrell Scholars Award.\\


\bibliographystyle{apj}
\small\bibliography{lit.bib}


\appendix
\section{APPENDIX A: Gound- Versus Space-Based Merger Grades}\label{sec:AA}

As mentioned, there exists a single 1200 sec {\it HST} exposure covering the inner $R_{\rm cl} \lesssim 500 \kpc$ of one of our clusters in {\it ACS} F606W.  We inspected the 80 ICBS targets (67, mass-complete sample) on this image for signs of major-mergers in a manner identical to that presented in Section 5 above. Reassuringly, although these data are of much higher spatial resolution, we find a merger-rate of $6\%\pm3\%$ (Poisson error), fully consistent with our ground-based estimate.

\begin{figure*}[h!]
\centering
\includegraphics[width = 0.7\linewidth]{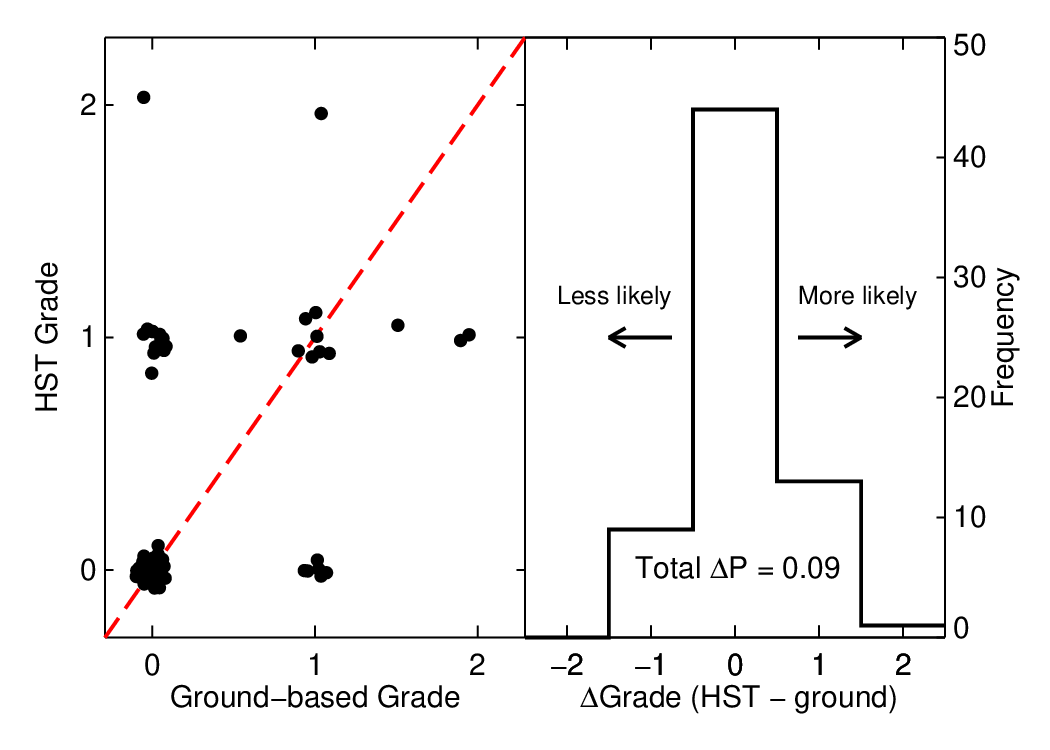}
\caption{Merger grades derived from {\it HST ACS} versus Magellan FourStar imaging.  {\it Left}: system-by-system correlation for galaxies in the mass-complete ICBS sample.  Points have been offset for clarity and half-grades come from averages over repeat assessments.  Scatter in this diagram is substantial, but no bias is apparent.  Thus, the ground-based merger-fractions discussed in Section 5 are, as argued, likely not to be underestimates.  Systems ``upgraded" in the {\it HST} data tend to have tidal/small-scale features unresolvable from the ground, but this enhanced resolution also separates blended galaxies, reducing their ground-based grades.  {\it Right}: Histogram of grading offsets.  Positive offsets indicate ``upgrades" using {\it HST} data.  Overall, these offsets amount to less than a $10\%$ increase in the probability that a galaxy would have received a higher merger grade if space-based data were available for all sources.  This is consistent with the upgrade rate derived from repeat inspections using the ground-based data and thus already accounted for in the estimates discussed in the text.}
\label{fig:HSTgrade}
\end{figure*}

Even in a system-by-system comparison, Figure \ref{fig:HSTgrade} (left) reveals that there is little-to-no bias in merger grades derived from these versus the ground-based data.  The scatter in this diagram is also revealing: { \it HST} ``upgrades" (points above the 1-to-1 dashed line) arise from the enhanced ability to resolve tidal features in the space-based data.  However, an almost equal number of systems become {\it downgraded} (pushed below the 1-to-1 line) since blending in the ground-based imaging is {\it also} resolved out.  Hence, apparently interacting galaxies in the ground-based data become clearly separated in the space-based imaging.

Due to this ``slosh", the integrated probability that a system would be upgraded using space-based data is only $+9\%$ (right panel), the same upgrade rate obtained from the ground-based inspection!  This good agreement gives us confidence that our merger estimate is, as argued, {\it not} biased low, further supporting our conclusion that major-mergers are not likely to be a large leak in our SB recycling scenario.

\end{document}